# Updated Results from VERITAS on the Crab Pulsar

**T. J. Nguyen for the VERITAS Collaboration**[*]

*Georgia Institute of Technology*
*E-mail:* tnguyen327@gatech.edu

The Crab pulsar and plerion are some of the brightest and best studied non-thermal astrophysical sources. The recent discovery of pulsed gamma-ray emission above 100 gigaelectronvolts (GeV) from the Crab pulsar with VERITAS (the Very Energetic Radiation Imaging Telescope Array System) challenges commonly accepted pulsar emission models and puts the gamma-ray emission region far out in the magnetosphere – close to or even beyond the light cylinder. We present updated VERITAS results from the analysis of a data set that is twice as large as the original data set published in 2011. The results are discussed in the context of discriminating between different models put forward to explain gamma-ray emission mechanisms and acceleration regions within the Crab pulsar's magnetosphere.

[*]Speaker

## 1. Introduction

The Crab pulsar, PSR B0531+21, is a young neutron star that powers the Crab Nebula, a remnant of the historical supernova observed in 1054 A.D. When the core of the progenitor star collapsed into a neutron star of 10 km radius, its magnetic flux and angular momentum was conserved. This gave rise to the pulsar's enormous surface magnetic field of 3.8 x $10^{12}$ G and a rotation period of ~33 ms [1]. Due to its proximity of ~2 kpc and intrinsic luminosity of 4.6 x $10^{38}$ erg $s^{-1}$ [1], the Crab is one of the best studied non-thermal astrophysical sources.

The search for pulsed emission from the Crab pulsar in very-high energy gamma rays has a long history. The Major Atmospheric gamma-ray Imaging Cherenkov (MAGIC) telescope was the first to detect pulsed emission above 25 GeV [2]. The detection favors the outer-gap model, which places the emission zone far out in the magnetosphere [2]. Some time later, observations with VERITAS revealed pulsed gamma-ray emissions at energies above 100 GeV, beyond the prediction from pulsar models at that time [3]. The detection placed significant constraints on the locations of gamma-ray emission and the emission mechanisms. In particular, the detection above 100 GeV challenges the curvature radiation mechanism and the geometry of the magnetic field [3]. According to the previous MAGIC and VERITAS measurements, the Crab pulsar's energy spectrum is harder than expected for an exponential cutoff at ~10 GeV [2-4], and therefore the exponential cutoff model was ruled out. However, more data is needed to distinguish between different pulsar models that differ in the spectral prediction above 400 GeV. Recently MAGIC has claimed a detection of pulsed emission from the Crab pulsar up to ~2 TeV with more than 300 hours of observations. This astonishing result needs verification given its implication on pulsar physics. In this paper, we present updated results from ongoing VERITAS observations of the Crab pulsar, with an emphasis on the pulsed emission at the highest energies.

## 2. Observation and Analysis

VERITAS, a major ground-based gamma-ray observatory located at the Fred Lawrence Whipple Observatory (FLWO) in southern Arizona (31 40N, 110 57W, 1.3 km a.s.l.), is an array of four imaging atmospheric Cherenkov telescopes designed to study gamma-ray astronomy in the very-high energy range [5]. Since the original VERITAS publication of the Crab pulsar observation in 2011, the number of hours of Crab pulsar observations has increased to a total of 194 hours. Most of the new data were collected after the camera upgrade in summer 2012, which resulted in an increase in VERITAS sensitivity and a lower threshold energy [6]. For this analysis, we selected data from observations carried out in wobble mode of 0.5 degree offset and zenith angles of 35 degree and below.

After data-quality selection, the recorded atmospheric shower images were processed using the data-analysis software package called VEGAS. We applied a standard principal component analysis to obtain the Hillas parameters, which parameterize the shower images [7]. The images from the four telescopes were then combined to reconstruct the events, i.e. the direction of origin and energy [8]. Monte Carlo simulations were used to optimize the event selection for the separation of gamma-ray events from cosmic-ray background events to obtain highest sensitivity in the pulsar analysis. The event selection parameters are the mean scaled width, mean scaled length, theta square (reconstructed squared angular distance to shower true direction), shower maximum height, and a lower size cut. The procedure to optimize the event selection assumed that the Crab pulsar energy spectrum followed a simple power law with an index of -4 and a gamma-ray flux of 1% of the Crab Nebula flux at 150 GeV.

For those events that survived the event selection, the arrival times were transformed to the barycenter of the solar system using the tempo2 pulsar timing package [9]. Using the ephemeris published monthly by the Jodrell Bank group for the Crab pulsar [10], the pulsar's rotational phase was calculated for each event. After phase folding, the pulse profile, or phaseogram, was constructed, as shown in Figure 1. To test the significance of the pulsed emission, we used the H-test [11] on the unbinned event times and the Li&Ma significance [12] calculated from the excess count of gamma rays at the main pulse (centered at phase 0.0 from Figure 1) and the interpulse (centered at phase 0.4 from Figure 1). The background region is indicated by the arrows and used to estimate the background contamination from cosmic rays and the Crab Nebula.

## 3. Results

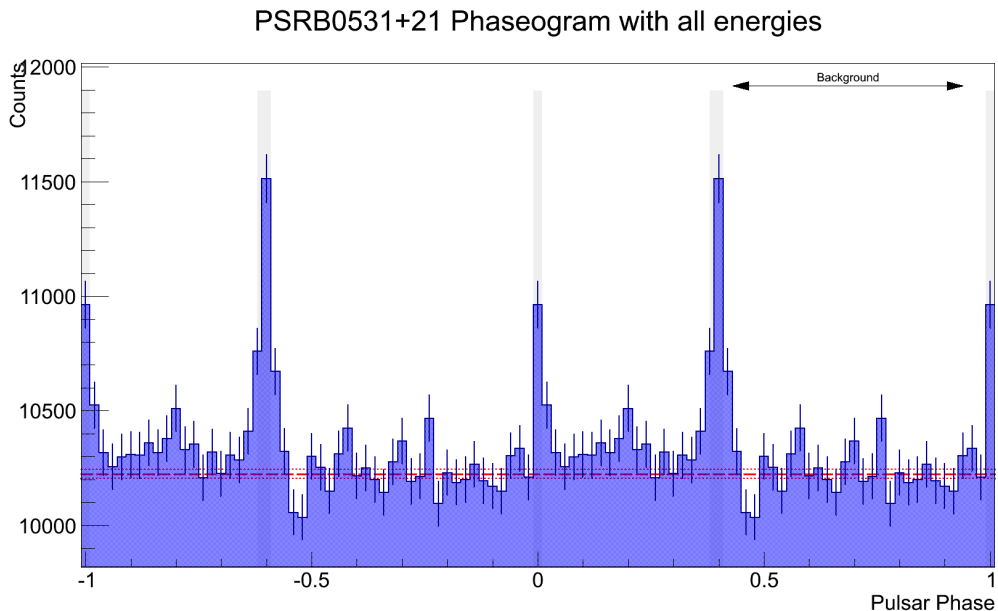

**Figure 1:** Phase-folded distribution of all energies. The two distinct pulses have peaks at phases 0.0 and 0.4 are called P1 and P2, respectively. The highlighted regions are the signal regions for P1 and P2, and the background is indicated by the arrows. The dashed, horizontal red line shows the background level estimated from the background region. The pulse profile is shown twice for clarity.

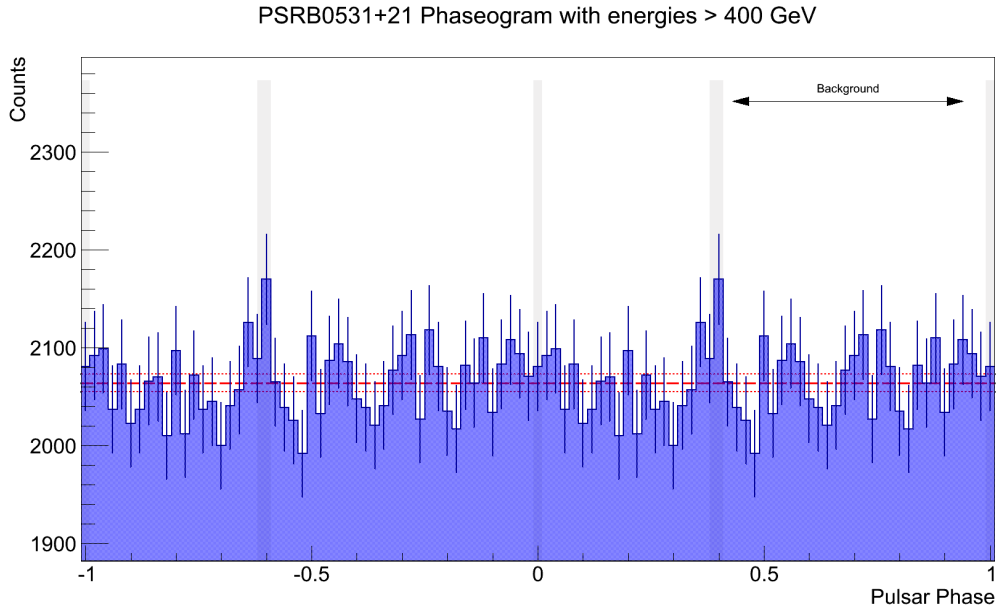

**Figure 2:** Phase-folded distribution with an energy threshold of E > 400 GeV.

The pulse profile of the accumulation of 194 hours of quality-selected data from observation of the Crab pulsar with VERITAS was constructed, as shown in Figure 1. The two distinct pulses occurring at phase 0.0 and phase 0.4 are the main pulse (P1) and the interpulse (P2), respectively. Previous published analysis showing narrower pulses at high energies [3] led to the selection of the signal regions to be between phases of -0.01 and 0.01 for P1 and between 0.38 and 0.41 for P2. The background region was chosen to be between phases 0.43 and 0.94. Events falling under these regions were selected to compute the pulsed excess count from both P1 and P2 combined, yielding a significance of 14.3σ.

Figure 2 shows the pulsar phaseogram composed of events with reconstructed energy above 400 GeV. The combined excess count from P1 and P2 only yielded 2σ. Thus a detection above 400 GeV cannot be claimed even with almost 200 hours of observation. There is, however, an indication that most of the signal is confined to P2 with no signal evidence at P1.

We reconstructed the differential energy spectrum of the Crab pulsar using the combined pulsed excess in the signal regions. The energy spectrum, shown in Figure 3 below, was well fitted with a power law $dN/dE = A(E/150 \text{ GeV})^{-\alpha}$, with the flux normalization $A = (3.6 \pm 0.6_{stat} + 2.4_{syst} - 1.4_{syst}) \times 10^{-7}$ TeV$^{-1}$ m$^{-2}$ s$^{-1}$, and the spectral index $\alpha = 3.5 \pm 0.5_{stat} \pm 0.2_{syst}$. Previous published measurement in 2011 (blue solid line) is also shown for the energy range from 100 GeV up to 400 GeV [3].

Curvature radiation originates from charged particles inside the pulsar's magnetosphere being confined to the magnetic field lines along a curved trajectories. These particles produce gamma rays with a natural break in energy spectrum at a few GeV within the radiation reaction limited regime [13]. The detection of pulsed emission above 100 GeV for the Crab pulsar suggested that the full picture of the radiation mechanism could not be explained by the curvature-radiation component alone at the very-high energy. In this work, we show an updated energy spectrum (red dashed line) that extends beyond 400 GeV and up to 1 TeV. The results agree well with the originally published VERITAS measurement (blue solid line). However, the accumulated statistics were not sufficient to reconstruct a spectral point at 1 TeV.

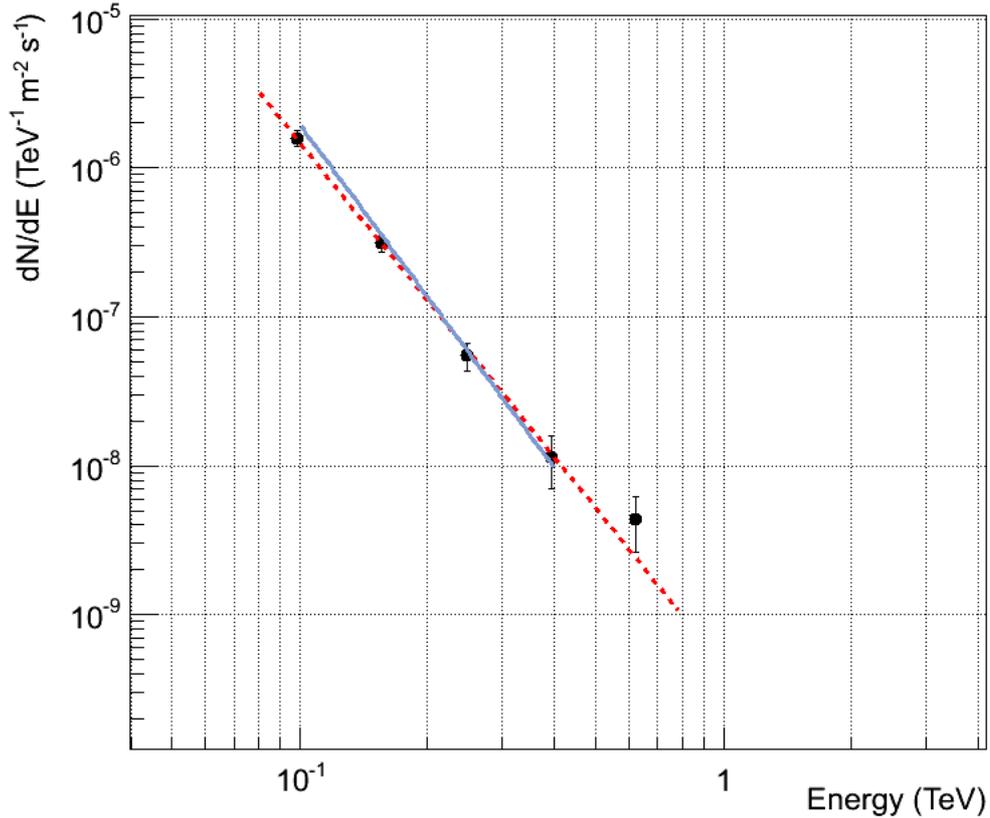

**Figure 3:** Phase-averaged differential energy spectrum of the Crab pulsar constructed from 194 hours of VERITAS observation. Also shown is the fitted spectral line (blue solid line) from VERITAS publication in 2011 [3].

## 4. Discussion and Conclusions

Although the Crab pulsar has been observed and studied extensively from radio to gamma-ray energies, the physics behind the emission mechanisms and the exact locations of particle acceleration regions within the pulsar magnetosphere still remain elusive. While there is an agreement that the acceleration zone is located in the outer magnetosphere, much debate still exists among different favored models for the radiation mechanisms within the magnetosphere [14]. Three major radiation mechanisms that are responsible for gamma-ray emissions are the synchrotron emission, inverse-Compton scattering, and curvature radiation.

So far the Crab pulsar is the only pulsar to display pulsed emission above 100 GeV [3], a detection that came as a surprise to the astronomical community. The detection has put strong constraints on the emission mechanisms and the acceleration regions within the magnetosphere. Currently, there are two possible explanations for the observation of pulsed emission at the very-high energy. It might be that a different mechanism other than curvature radiation is responsible for all pulsed emissions, or that a second mechanism becomes more dominant above the spectral break [3]. Our updated analysis of the Crab pulsar's energy spectrum further supported such view. However, more data is needed to make a definite conclusion about the spectral shape above 400 GeV. With 194 hours of observations, we are unable to confirm the MAGIC detection of pulsed emission up to ~2 TeV. However, the Crab observation with VERITAS is currently planned to extend up to 300 hours, which will enable us to weigh in on the MAGIC claim with the same or even more sensitivity.

**Acknowledgments**

This research is supported by grants from the U.S. Department of Energy Office of Science, the U.S. National Science Foundation and the Smithsonian Institution, and by NSERC in Canada. We acknowledge the excellent work of the technical support staff at the Fred Lawrence Whipple Observatory and at the collaborating institutions in the construction and operation of the instrument. The VERITAS Collaboration is grateful to Trevor Weekes for his seminal contributions and leadership in the field of the VHE gamma-ray astrophysics, which made this study possible.